\documentclass[11pt]{openjournal}
\usepackage{tabularx}
\usepackage{graphicx}
\usepackage{cancel}
\usepackage{multirow}
\usepackage{amssymb}

\usepackage{amsmath}

\usepackage{color}
\usepackage[pdftex, colorlinks=true, linkcolor=black, urlcolor=black,  citecolor=blue]{hyperref}
\usepackage{subfigure} 
\usepackage{float}

\begin{document}
\title{Classifying Exoplanets with Gaussian Mixture Model}

\author{Soham Kulkarni$^1$}
\author{Shantanu Desai$^2$}
\address{$^1$Dept. of Physics, University of Florida, Gainesville, Florida, 32611, USA}

\address{$^2$Dept. of Physics, Indian Institute of Technology, Hyderabad,
Kandi, Telangana 502285, India}

\begin{abstract}
\noindent
Recently,  Odrzywolek and  Rafelski~\cite{exoplanetclass} have found three distinct categories of exoplanets,
when they are classified based on density.
We first carry out  a similar  classification of exoplanets according to  their density  using the Gaussian  Mixture Model, followed by information theoretic criterion (AIC and BIC) to determine the optimum number of components. Such a one-dimensional classification favors two components using AIC and three using BIC, but the statistical significance from both the tests is not significant enough to decisively pick the best model between two and three components.  We then extend this  GMM-based classification to two dimensions by using both the density and the Earth similarity index~\citep{Kashyap}, which is a measure of how similar each planet is compared to the Earth.  For this two-dimensional classification, both  AIC and BIC provide  decisive  evidence in favor of three components. 
\end{abstract}

\maketitle

\section{Introduction}
\label{sec:intro}
\noindent

Over the past two decades, there has been a revolution in the field of exoplanet astronomy following  the confirmation of more than 3000 planets orbiting stars other than the Sun (see Ref.~\cite{Rice} for a recent review and Ref.~\cite{Wei18} for a summary of exoplanet detection techniques). Lot of work has been done to characterize the properties of the exoplanets discovered using the myriad techniques~\cite{lunine}. Recently, Odrzywolek and Rafelski~\cite{exoplanetclass} (hereafter OR16) have carried out  the classification of exoplanets according to their density, following a suggestion long time back~\cite{Weisskopf}. 
OR16 fitted  the exoplanet density data  to  lognormal distributions to determine the optimum number of components. They found three lognormal components with peak densities at 0.71 $~\rm{gm/cm^3}$, 6.9 $~\rm{gm/cm^3}$, and 29.1 $~\rm{gm/cm^3}$~\cite{exoplanetclass}. These  three components correspond to ice/gas giants, iron/rock super-Earths, and brown dwarfs  respectively.  The optimum number of components was determined by maximizing the log-likelihood and then checking the goodness of fit for different number of components by calculating the $p$-value from three distinct non-parametric tests.

We would like to do a variant of the above analysis by carrying out a similar classification according to density using Gaussian mixture models, followed by   information theoretic criterion to determine the optimum number of classes.  We have previously  used this procedure, to perform a  unified classification of all GRB datasets using three different model comparison techniques~\cite{Kulkarni}. We then extend the analysis of OR16 by considering two-dimensional classification using both the density and Earth similarity index.

This paper is organized as follows. In Section~\ref{sec:data}, we describe the dataset and the physical quantities used for the classification. The mathematical basis for the classification is discussed in Section~\ref{sec:analysis}. Our results are shown in Section~\ref{sec:results} and we conclude in Section~\ref{sec:conclusions}.

\section{Data}
\label{sec:data}

\subsection{Exoplanet Catalog}
\noindent

In the one-dimensional classification, we shall  study the trends obtained from  the confirmed detections, by classifying the exoplanet database according to their densities, for which we need the mass and radius of the planets. We obtain the mass and radius information from the catalogs uploaded on the  NASA Exoplanet archive\footnote{\url{https://exoplanetarchive.ipac.caltech.edu/}} and the Extrasolar planet encyclopedia\footnote{\url{http://exoplanet.eu}}  as of \textbf{February 18, 2017}. From these datasets, we consider only those planets with measured values of mass and density, and which exist in both the datasets with the same observed values to avoid any irregularities and to maintain consistency in the dataset. The NASA Exoplanet archive is a NASA funded public data service, which is hosted by the  Infrared Processing and Analysis Center.  This catalog lists only  those objects, for which their detection and planetary status is sacrosanct. As of Feb 18, 2017, it   contained a total of 3440 planets out of which 531 have  measured mass and radius values detected. Most of the planets listed in this catalog have been detected using transit photometry.
The Extrasolar planet encyclopedia is maintained by the Meudon Observatory in Paris and as of Feb 18, 2017 contained  total of 3567  planets (most of which were also detected using transit photometry), of which 615 have  measured values for  all the parameters. The data provided by the two catalogs is similar except for some differences in their selection criteria. The Extrasolar planet encyclopedia allows planets weighing from 60 Jupiter Mass onwards, whereas the NASA Exoplanet archive uses 30 Jupiter mass as the lower limit, which is also the reason for the smaller  number of confirmed exoplanets in the latter. However, one caveat is that the catalog is continuously updated and sometimes false detections are removed as the data gets subjected to more scrutiny. Therefore,  in order to obtain a gold sample, we  have selected 450 observations, which are common to both the datasets for our study in this paper.
Both the datasets used for this analysis as well as the code which looks for common planets between the two catalogs have been uploaded on {\tt github} and can be found at \url{https://github.com/IITH/Exoplanet-Classification} 
\\

In addition to  the one-dimensional classification using only density, we also carry out a two-dimensional classification, wherein we  use both the density and the  Earth Similarity Index (or ESI)~\cite{Kashyap} for the classification. For this,  we need some additional parameters for the calculation of ESI. The additional parameters that we need apart from the radius and density are the surface temperature and time period of revolution, as other parameters can be derived from the mass and radius. The escape velocity and surface gravity are calculated by positing that the shape of the planet is a perfect sphere, wherein the total mass is distributed uniformly throughout the volume. We only consider planets for which we have  the observed values for  all four of these parameters.

\subsection{Calculations for the data:} 
Assuming the planet is a perfect sphere with a  uniform mass distribution, the expression for density is: 
\begin{equation}
    \bar{\rho} = \frac{M}{\frac{4}{3}\pi R^{3}}
\end{equation}

The escape velocity is given by: 

\begin{equation}
    v_{esc} = \sqrt{\frac{2GM}{R}}.
\end{equation}

The surface gravity is obtained from:

\begin{equation}
    g_{surf} = \frac{GM}{R^2}.
\end{equation}

where $G$ is the Gravitational Constant, $M$ is the mass of the planet and $R$ is the radius. \\ 


ESI is a figure of merit used to ascertain how habitable is the planet for life to develop compared to the Earth.  More details on the theory behind ESI can be found in the work by Kashyap~\cite{Kashyap}, which in turn  follows the prescription  from Schulze-Makuch et al.~\cite{Schulze} (See also ~\cite{Moya} for alternate indices proposed similar in spirit to ESI).  The ESI is  based on six different parameters, viz. density, radius, temperature, surface gravity, escape velocity, and the time period of revolution around their Sun. 
All these parameters are normalized to  Earth units, as it is convenient for the index calculation.  The ESI is calculated based on the Bray-Curtis Similarity index~\cite{Bray} and is given by:

\begin{equation}
ESI_{x} = \left(1- \left|\frac{x-x_0}{x+x_0}\right|\right)^w
\label{eq:ESI}
\end{equation}

\noindent where $x$ is the parameter for which the index has to be calculated, $x_0$ is the reference values which in our case is one, as we have expressed all parameters in Earth units and $w$ is the weight exponent. 

The total ESI is given by:
\begin{equation}
ESI = \left(ESI_{g}\times ESI_{temp}\times ESI_{vesc}\times ESI_{p}\times ESI_{r}\times ESI_{d}\right)^{1/6}
\end{equation}
The values of ESI range from 0 (completely different from Earth) to 1 (resembling  a clone of Earth).

\section{Analysis Methods:}
\label{sec:analysis}
\noindent
We outline the method used for both the one-dimensional classification using density and the two-dimensional classification using density and ESI.
For finding the best-fit parameters, we use the Gaussian-mixture Model (GMM)~\cite{astroml}, which is part of the {\tt Scikit-learn} package, used for a variety of machine learning applications in python. The GMM fits the data to a mixture of multiple ($k$) lognormal Gaussian distributions, which are characterized by their mean, covariance and their respective weights in the fit data. The GMM method uses the Expectation Maximization (EM) algorithm~\cite{EM} to maximize the likelihood function over the given parameter space. The GMM method can also be generalized to include error bars and this generalized GMM algorithm is referred to in the astrophysics literature as Extreme Deconvolution~\cite{ED}.
However, since we are using a planet catalog measured in two separate datasets having negligible error bars, we stick to the ordinary GMM method.
Given the probability distribution  function $f(x,\theta)$, where $x$ are the observed datapoints, $\theta$ are the parameters used to define the function, $N$ being the total number of exoplanets in our study, and $w_k$ denotes the weights associated with each of the $k$ log normal distributions, the likelihood can be defined as:

\begin{equation}
    \mathcal{L} = \sum\limits_{i=1}^{N} \sum_{j=1}^{k} w_{j}f_{j}(x_{i},\theta),
    \label{eq:likelihood}
\end{equation}
and the probability distribution function for a univariate Gaussian as:

\begin{equation}
    f(x,\theta) = \frac{1}{\sqrt{2\pi} \sigma} \exp\left(- \frac{(x-\log \rho_{planet})^{2}}{2\sigma^{2}}\right).
\end{equation}
 A generalized  bivariate Gaussian distribution can be defined as:

\begin{equation}
    f(x,\theta) = \frac{1}{2\pi \sigma_1 \sigma_2 \sqrt{1-\rho'^2}} \exp\left[- \frac{1}{2(1-\rho'^2)}\left( \frac{(x-\mu_{x})^{2}}{\sigma_1^2} + \frac{2\rho (x-\mu_{x})(y-\mu_y)}{\sigma_1 \sigma_2} + \frac{(y-\mu_y)^{2}}{\sigma_2^2}\right)\right]
\label{eq:2dfit}    
\end{equation}

\noindent where $\rho' = \frac{V_{12}}{\sigma_1 \sigma_2}$ is the correlation, $V$ is the covariance of the two variables and $\mu_x$ is the mean log(density) and $\mu_y$ is the mean ESI. An additional condition being used in the EM algorithm is the normalization condition:

\begin{equation}
    \sum\limits^{k}_{i=1} w_{i} = 1
\end{equation}

In this study, we use the GMM method for the $k=2$ and $k=3$ lognormal fits to the data followed by information theory based model comparison methods to assess the best fit amongst these two models.

\subsection{Model Comparison}

Once we have obtained the best-fit parameters for each model, we need to select the optimum model from all the possibilities  being considered. Naively, the simplest way to do model comparison would be   by carrying out likelihood comparison between the competing models and choosing the model with the highest likelihood as the best model. However, the maximization of likelihood could lead to an overfitting  of the model to the data with additional degrees of freedom and hence we need a more robust and accurate criterion, which will penalize the use of extra free parameters. This can be done by using the Information criterion tests, such as  Akaike Information Criteria (AIC) and the Bayesian Information Criteria (BIC), which are commonly used in Astrophysics literature~\cite{Shi,Shafer,Desai16a,Ganguly,Liddle} (and references therein). These information criteria-based methods provide a way to penalize  the excess free parameters and determine the best model accordingly.  

\subsubsection{AIC:}
\label{sec:aic}
Akaike Information Criteria or AIC~\cite{Burnham} penalizes lightly the excess free parameters  and is defined as: 

\begin{equation}
AIC = 2p + 2 \ln L
\label{eq:aic}
\end{equation}

\noindent where $p$ is the number of free parameters in the model and $L$ is the likelihood.
The AIC defined in Eq.~\ref{eq:aic} is valid when the ratio $N/p$ is very large i.e. $>40$. For a ratio less than this, a first order correction is included and the modified expression is given by:

\begin{equation}
AIC = 2p + 2 \ln L + \frac{2p(p+1)}{N-p-1}
\end{equation}

\noindent Throughout  our data, the ratio is greater than the value prescribed and hence we do not account for this correction in our study. The preferred model is the one with a lower value of AIC and the efficacy of this hypothesis is determined using the quantity: 

\begin{equation}
\Delta AIC_i = AIC_{i} - AIC_{min},
\end{equation}

\noindent where  $\Delta AIC_i$ value corresponds to the preference of the model $i$ over the model with the lower AIC value and hence is the null hypothesis. The confidence in the model can be determined by the magnitude of the $\Delta AIC$ value. Although one cannot formally calculate $p$-values from $\Delta AIC$, one usually uses qualitative strength of evidence rules to judge the efficacy of a given model~\cite{Shi,Liddle,Liddle07}.
As pointed out by  Liddle~\cite{Liddle}, the value for the best model will be, $\Delta AIC_i = 0$. Now, if $0 < \Delta AIC_i < 2$, then we can say that we have a weak or no statistical evidence to reject the $i^{th}$ model over the null hypothesis.
$2 < \Delta AIC_i < 6$ implies that the model has only weak support and has evidence against this model. For models with $\Delta AIC_i > 6$ there exists strong evidence against the model and $\Delta AIC_i > 10$ implies a very strong  or decisive evidence against the $i^{th}$ model. These rules  can be applied directly for the $BIC$ criterion (next subsection) as well.

\subsubsection{BIC:}
Bayesian Information Criterion or BIC was used by Schwarz~\cite{Schwarz} and is used to penalize the free parameters much more harshly than the AIC criterion and is defined as:

\begin{equation}
BIC = p \ln N + 2 \ln L
\end{equation} 
\noindent Again, the preferred model is the one with the lower values of BIC and is taken as the null hypothesis for further determining the significance of different models.

\begin{equation}
\Delta BIC_i = BIC_{i} - BIC_{min}
\end{equation}

Similar to the significance test for the AIC criterion, the $\Delta BIC_i$ value acts as the significance measure for the BIC test and follows the same values as for AIC. The only difference being that according to  BIC criterion, the penalty for a model with extra number of free parameters is harsher  compared to  AIC.  


\section{Results:}
\label{sec:results}
\subsection{1D classification}
We first describe our results for the one-dimensional classification using only the density.
We apply the techniques and methods described in the previous sections to the exoplanet catalog, generated by filtering the data from the NASA Exoplanet archive and the Extrasolar Planet encyclopedia as mentioned earlier. For the density function, we find the best-fit model parameters for $k$ lognormal distributions according to the density from Eq.~\ref{eq:likelihood}. Each distribution is characterized by its mean, standard deviation and the weight of the distribution indicating the number of planets that have been classified under that particular distribution. We apply  the GMM routine to the density functions after varying the  number of Gaussians from 1 to 14.
\begin{figure}[h!]
\centering
\includegraphics[scale = 0.5]{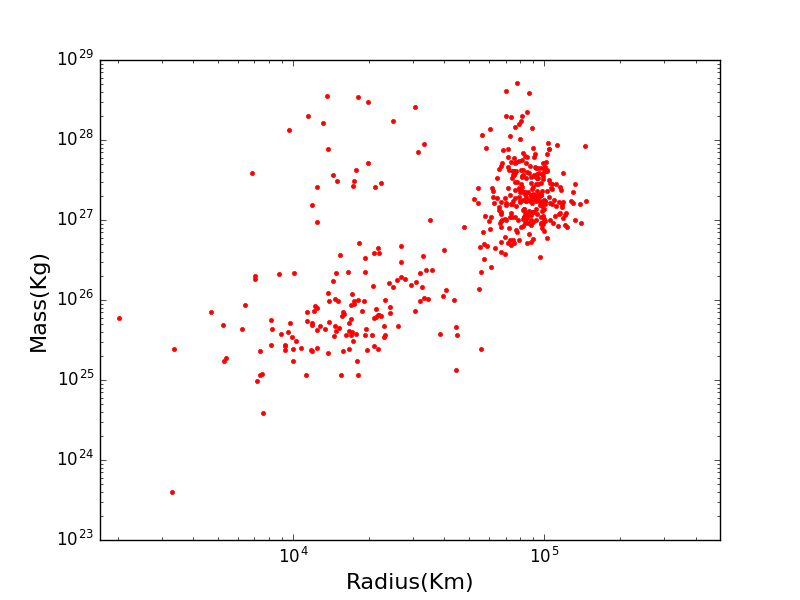}
\caption{A scatter plot of the mass versus radius for the sample dataset encompassing all the 450 planets used in the study.}
\label{fig:fig1}
\end{figure}

\begin{figure}[h!]
\centering
\includegraphics[scale=0.5]{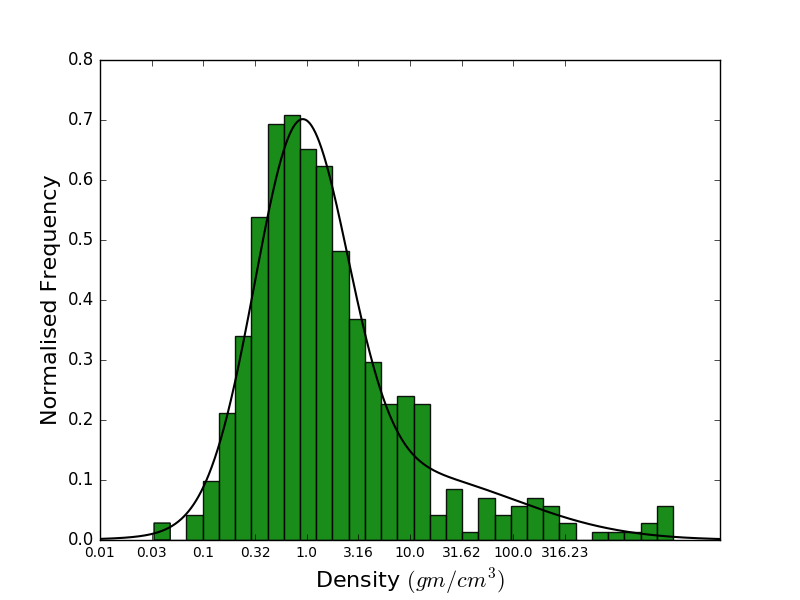}
\caption{The  GMM based fit  for the density of the exoplanets using the best-fit parameters from Eq.~\ref{eq:likelihood} for $k=2$. Details of the fit can be found in Tab.~\ref{tab:aicbic}.}
\label{fig:2ghist}
\end{figure}

\begin{figure}[h!]
\centering
\includegraphics[scale=0.5]{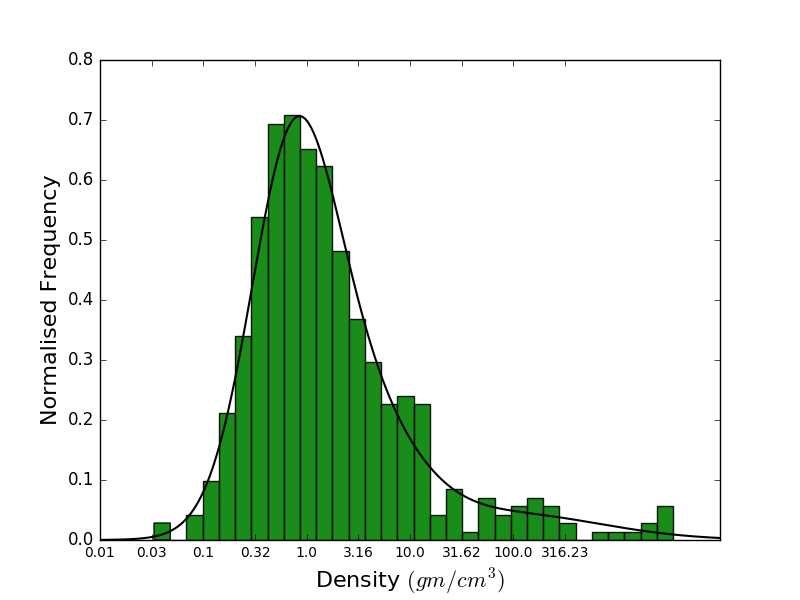}
\caption{The GMM based fit  for the density of the exoplanets using the best-fit parameters from Eq.~\ref{eq:likelihood} for $k=3$. Details of the fit can be found in Table~\ref{tab:aicbic}.}
\label{fig:3ghist}
\end{figure}

\begin{figure}[h!]
\centering
\includegraphics[scale=0.5]{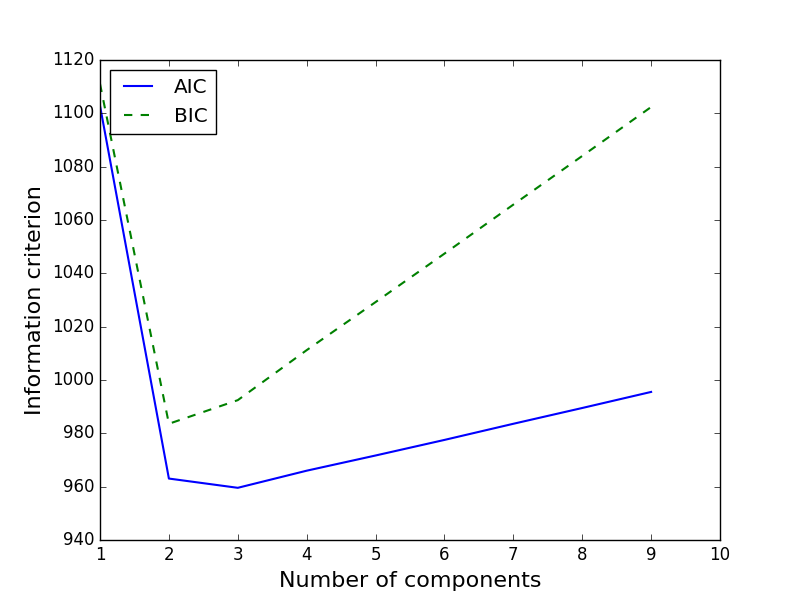}
\caption{The AIC and BIC values as a function of the number of Gaussian components used to fit the density of exoplanets. AIC shows a preference for two components, whereas BIC shows a preference for three.}
\label{fig:aicbic}
\end{figure}

\begin{table*}
\centering
\caption{Model Comparison parameters for exoplanets, when classified according to density. The first  column denotes
 the number of components used for the fitting.
The second column is the best-fit value for the logarithm  of the density (expressed in $~\rm{gm/cm^3}$), the third column is its standard deviation, and the fourth column indicates the number of planets classified in each category. The next two columns indicate the AIC and BIC value for both the hypothesis. The last two columns indicate the absolute value of $\Delta$ (AIC/BIC) between the model with minimum value and the next highest value. As we can see, AIC prefers 3 components, whereas BIC prefers 2. However, $\Delta$AIC and $\Delta$BIC are both $<10$ and hence  their significance (compared to the value with higher AIC/BIC) is marginal.}
\begin{tabular}{|c|ccc|cc|cc|}
\hline
$k$ & $ \mu$ & $\sigma $ & $w_{i}$ & $AIC$ & $BIC$ & $\Delta(AIC) $& $ \Delta(BIC)$ \\
\hline
\multirow{2}{*}{2} & 0.88 & 0.20 & 322 &  \multirow{2}{*}{965.7} & \multirow{2}{*}{\textbf{986.2}}& \multirow{5}{3em}{5.6} & \multirow{5}{3em}{6.6} \\
\cline{2-4}
& 9.69 & 1.08 & 128 & & & &\\
\cline{1-6}
\multirow{3}{*}{3} & 0.71 & 0.17 & 225 & \multirow{3}{*}{\textbf{959.6}} & \multirow{3}{*}{992.8} & & \\
\cline{2-4}
& 2.03 & 0.36 & 175 & & & &\\
\cline{2-4}
& 88.0 & 0.82 & 50 & & & &\\
\hline
\end{tabular}
\tablecomments{In the table the preferred value for every test is highlighted in bold.}
\label{tab:aicbic}
\end{table*}

The scatter plot in Fig.~\ref{fig:fig1} shows all the selected planets for the study  as a function of their mass and radius. The distribution looks clustered in certain areas with  lots of outliers.
The density distribution of 450 exoplanets with their histograms can be found in Fig.~\ref{fig:2ghist} and Fig.~\ref{fig:3ghist} for the 2-Gaussian and 3-Gaussian fits respectively and we can see intuitively that no difference can be discerned by eye  from the two figures. Both the models fit well the distribution of the density function, and hence we have to rely on  quantitative model comparison tests that have been carried out on the data, viz. the AIC and the BIC test. As seen in Fig.~\ref{fig:aicbic}, the BIC test indicates that  the 2-component model is  the optimum  model as it has the minimum BIC value followed by the 3-component model, which has a larger value than the two component model. This trend is different from the AIC test, as the AIC has a minimum for  three Gaussians, indicating that this is the best model, followed by the two-component model. 
These results if compared to the previous attempts at one-dimensional classification done by OR16 are very similar in both, the two Gaussian and the three Gaussian models  proposed  in this paper. From the  2-component model,   the mean density values are  at $0.88 ~\rm{gm/cm^3}$ and $9.69 ~\rm{gm/cm^3}$, with each class containing 322 and 128 exoplanets respectively. The 
inferred mean values of the density for the 3-component model are  at $0.71 ~\rm{gm/cm^3}$, $2.03 ~\rm{gm/cm^3}$ and $88.1 ~\rm{gm/cm^3}$ with 225, 175, and 50 in each of the classes respectively. In the previous study by OR16,  the mean density values  are at $0.7 ~\rm{gm/cm^3}$ and $6.3 ~\rm{gm/cm^3} $ with 320 and 106 respectively in each class for 2 components and at $0.71 ~\rm{gm/cm^3}$, $6.9 ~\rm{gm/cm^3}$ and $29.1 ~\rm{gm/cm^3}$ with 340, 80, and 7 exoplanets respectively in each class for the 3 components.  

From  Tab.~\ref{tab:aicbic}, we see that for the AIC test, the best model preferred is the 3-Gaussian model but there is sufficient confidence shown in both, the 2-Gaussian model and surprisingly the 4-Gaussian model (see Fig.~\ref{fig:aicbic}) while rejecting all other models by a huge margin. As described in Sect.~\ref{sec:aic}, the intervals of $\Delta_i$ are well within the range of not having sufficient evidence to reject the 2-component and 4-component models over our null hypothesis of a 3-component model. The BIC test prefers the 2-component model and has weak confidence in the 3-Gaussian model while rejecting all other models by a significant margin and hence rejecting the 4-component model as well from further consideration. Therefore, the results from the two information criterion tests do not agree. However, $\Delta$AIC and $\Delta$BIC are both less than 10 between the two and three component model, so no one model among these is decisively favored between the two.

\subsection{2D Classification:}
We now proceed to a  2-dimensional GMM based classification using both the logarithm of the density and ESI.
We use the combined data from  ESI and density  using the datasets  specified earlier in the manuscript and perform a two-dimensional GMM analysis. We consider only the planets that have measured values for all the quantities required for the calculation of ESI. A total of 450 exoplanets were analyzed for a range of lognormal components. 

\begin{figure}[h!]
\centering
\includegraphics[scale=0.38]{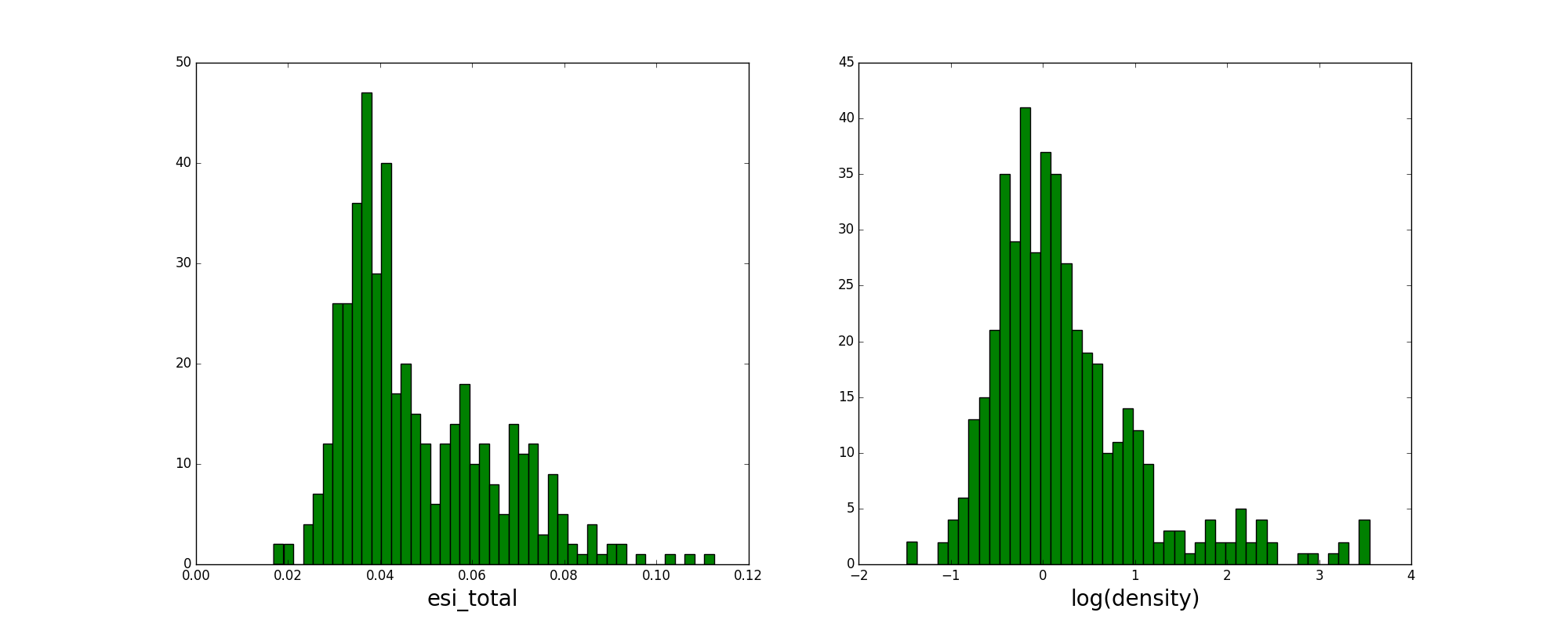}
\caption{The histograms for the density (expressed in $\rm{gm/cm^3}$) and the total ESI shown individually to see the general trend of the distribution.} 

\end{figure}

\begin{figure}[h!]
\centering
\includegraphics[scale=0.35]{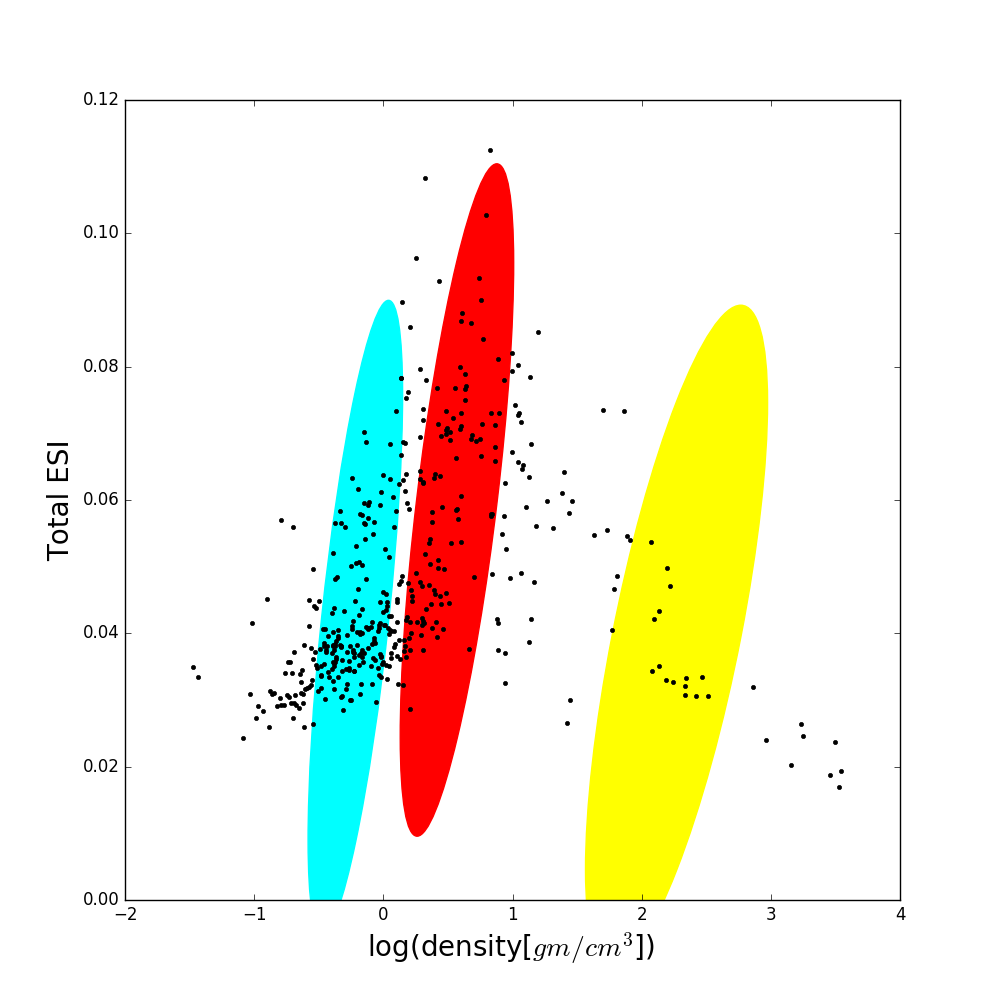}
\caption{The scatter plot of the distribution using the two components, log (density) and total ESI. The three  ellipses  represent  the $1\sigma$ confidence level region for the 3-component model,  which are centered at the means of the distribution acquired from best-fit of Eq.~\ref{eq:likelihood} and Eq.~\ref{eq:2dfit}.}

\label{fig:denesi}
\end{figure}

\begin{figure}[h!]
\centering
\includegraphics[scale=0.5]{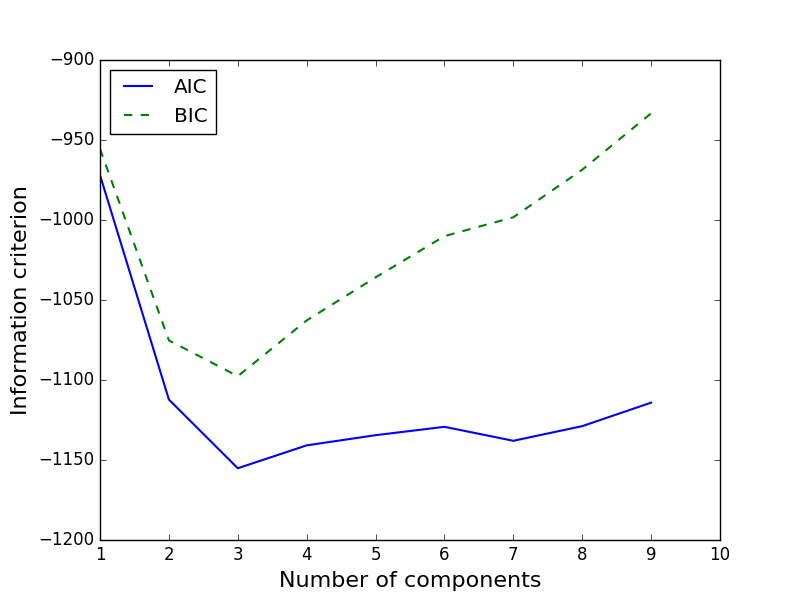}
\caption{AIC and BIC values for the two dimensional GMM analysis (as a function of log(density) and ESI) over the combined data. Both AIC and BIC attain a minimum value for three components. }
\label{fig:denesiaic}
\end{figure}

As we can see from Fig.~\ref{fig:denesiaic}, we get a result that is similar to the one we saw in the above case of 1-D classification, where the 3-component component was preferred  only with AIC, albeit with marginal significance, using only the density as a parameter.
From this two-dimensional analysis using the total ESI along with the density, we have both AIC and BIC preferring the 3-component distribution over all the other ones and by a substantial margin. The best-fit values of the parameters along with their covariance, as well as the  $\Delta AIC$ and $\Delta BIC$ values for the  two and three component distributions can be found  in Tab.~\ref{tab:aicbic2d}


\begin{table*}[!htbp]
\caption{Results from model comparison  for 2D classification of exoplanets using logarithm of the density and ESI, along with results from information theoretic criterion. The explanation for the  first two  columns is same as in Tab.~\ref{tab:aicbic}. The third column denotes the covariance matrix. Other columns are same as in Table~\ref{tab:aicbic}. For this case, both AIC and BIC point to decisive evidence for three components.}
\label{tab:aicbic2d}
\begin{tabular}{|c|ccc|cc|cc|}
\hline
$k$ & $\mu$ & $\Sigma$ & $w_{i}$ &   AIC  &  BIC   & $\Delta(AIC) $& $ \Delta(BIC)$ \\
\hline
\multirow{2}{*}{2} & (-0.063,  0.046) &  $\left(\begin{array}{cc}                            0.24 &0.005\\0.005 &0.0012 \\               \end{array}\right)$ & 332       & \multirow{2}{*}{\textbf{-1112.2}} & \multirow{2}{*}{\textbf{1075.2}} & \multirow{5}{*}{43.0} & \multirow{5}{*}{22.4} \\
\cline{2-4}
& (1.07,0.052) &$\left(\begin{array}{cc}       0.902 &-0.01\\ -0.01 &0.0013 \\     \end{array}\right)$ & 118 & &  & &\\
\cline{1-6}
\multirow{3}{*}{3} &(-0.22,0.042) &$\left(\begin{array}{cc}       0.17 &0.0036\\ 0.0036 & 0.012 \\     \end{array}\right)$ & 270     & \multirow{3}{*}{-1155.2} & \multirow{3}{*}{-1097.7}  & & \\
\cline{2-4}
& (0.57,0.06) &$\left(\begin{array}{cc}       0.33 &0.0056\\ 0.0056& 0.0013 \\     \end{array}\right)$ & 143 & &  & & \\
\cline{2-4}
& (2.27,0.04) &$\left(\begin{array}{cc}       0.68 &-0.011\\ -0.011 & 0.0013 \\     \end{array}\right)$& 37 & &  & & \\
\hline
\end{tabular}
\end{table*}

The AIC and BIC tests both point to definitive evidence for one model (three components) and give concordant results. From the statistical confidence measures $\Delta AIC $ and $\Delta BIC$, we can assert our confidence in the hypothesis of the three-Gaussian model over all  other model fits. We find  that for the next preferred models in  the analysis, the $\Delta AIC = 14 $ and $\Delta BIC = 22$, which is significant enough to reject the respective models in favor of our null hypothesis with strong confidence.

\section{Conclusions}
\label{sec:conclusions}
In this manuscript, we have undertaken a classification of the exoplanet catalog using clustering based on the logarithm of the planet  density (similar to a recent analysis in OR16~\cite{exoplanetclass}), followed by   a 2-dimensional analysis using both the log of density and Earth Similarity Index (ESI)~\cite{Kashyap} for  each of the exoplanets. We use Gaussian Mixture Model to classify the data  for both the one-dimensional and two-dimensional classifications based on log(density) and \{log(density), ESI\} respectively. For both of these classifications, we determine the best-fit parameters for each model using the EM algorithm. We then use information theoretic criterion, such as AIC and BIC  to determine the optimum number of free parameters. Our results are as follows:

\begin{enumerate}
\item For the  one-dimensional approach, our analysis does not provide  a conclusive evidence  between a two-component and a three-component model, since neither of the information criterion tests cross the threshold ($>10$) needed for decisive evidence. As  stated in Tab.~\ref{tab:aicbic}, the $\Delta AIC$ test weakly favors the three component Gaussian model, whereas the $\Delta BIC$ test weakly favors the two component Gaussian model. The 2 Gaussian model has the mean values of the density at $0.88 ~\rm{gm/cm^3}$ and $9.69 ~\rm{gm/cm^3}$, whereas the corresponding values for  the  3 Gaussian model are located at $0.71 ~\rm{gm/cm^3}$, $2.03 ~\rm{gm/cm^3}$ and $88.1 ~\rm{gm/cm^3}$. 

\item The two-dimensional classification on the other hand provide robust and consistent results from both the tests. As is summarized in Tab.~\ref{tab:aicbic2d}, both the tests give decisive evidence for the  three component Gaussian model with $\Delta$AIC and $\Delta$BIC $> 10$ in both the cases.

The catalogs used for this analysis (which were downloaded on Feb 18, 2017) along with the code used can be found online at \url{https://github.com/IITH/Exoplanet-Classification}.

\end{enumerate}

 \noindent

\bibliographystyle{utcaps}
\bibliography{exoplanet}

\end{document}